\documentclass[prc,twocolumn,showkeywords,showpreprintnumbers,floatfix,nofootinbib]{revtex4-1}
\usepackage{amsmath,amssymb,graphicx,bm}
\newcommand{\vek}[1]{\bm{\mathrm{#1}}}

\DeclareMathOperator{\im}{Im}
\DeclareMathOperator{\re}{Re}
\newcommand{\nablav}{\vek{\nabla}}
\newcommand{\laplace}{\Delta}
\newcommand{\pv}{\vek{p}}
\newcommand{\kv}{\vek{k}}
\newcommand{\Kv}{\vek{K}}
\newcommand{\qv}{\vek{q}}

\newcommand{\jv}{\vek{j}}
\newcommand{\Jv}{\vek{J}}
\newcommand{\sv}{\vek{s}}
\newcommand{\Tv}{\vek{T}}
\newcommand{\sigv}{\vek{\sigma}}
\newcommand{\JJ}{\mathbb{J}}
\newcommand{\qtil}{\tilde{q}}
\newcommand{\ttil}{\tilde{t}}
\newcommand{\Pitil}{\tilde{\Pi}}
\newcommand{\vlowk}{\ensuremath{V_{\text{low}\,k}}}
\newcommand{\ph}{\text{ph}}
\newcommand{\RPA}{\text{RPA}}
\newcommand{\fm}{\,\text{fm}}
\newcommand{\fmi}{\,\text{fm}^{-1}}
\newcommand{\MeV}{\,\text{MeV}}
\newcommand{\dotp}{\mkern -2mu \cdot\mkern -2mu}
\newcommand{\crossp}{\mkern -4mu \times\mkern -4mu}
\newcommand{\CG}[6]{C_{#1 #2 #3 #4}^{#5 #6}}
\newcommand{\half}{\frac{1}{2}}
\newcommand{\Sec}[1]{Sec.~\ref{#1}}

\newcommand{\Ref}[1]{Ref.~\cite{#1}}
\newcommand{\Refs}[1]{Refs.~\cite{#1}}
\newcommand{\Eq}[1]{Eq.~(\ref{#1})}
\newcommand{\Eqs}[1]{Eqs.~(\ref{#1})}
\newcommand{\Fig}[1]{Fig.~\ref{#1}}

\newcommand{\calV}{\mathcal{V}}
\newcommand{\tot}{\text{tot}}
\newcommand{\corr}{\text{corr}}
\newcommand{\Gtemp}{\mathcal{G}}

\begin{document}

\title{Neutron pairing with medium polarization beyond the Landau
  approximation} \author{M. Urban} \email{urban@ipno.in2p3.fr}
\affiliation{Institut de Physique Nucl\'eaire, CNRS-IN2P3,
  Univ. Paris-Sud, Universit\'e Paris-Saclay, 91406 Orsay cedex,
  France} \author{S. Ramanan} \email{suna@physics.iitm.ac.in}
\affiliation{Department of Physics, Indian Institute of Technology
  Madras, Chennai - 600036, India}
  
\begin{abstract} We revisit the long-standing problem of the superfluid
  transition temperature $T_c$ in dilute neutron matter. It is well
  known that $T_c$ is strongly affected by medium polarization effects
  (screening) which modify the pairing interaction in the medium. We
  study these effects within the random-phase approximation (RPA). It
  turns out that the widely used Landau approximation is sufficient
  only at densities below about $0.002\fm^{-3}$. At higher densities,
  the full RPA leads to stronger screening than the Landau
  approximation.
\end{abstract}
\maketitle
\section{Introduction}
Superfluidity in neutron stars is a long-standing problem that finds
its inception in the attractive interaction between two nucleons that
allows for the formation of Cooper pairs~\cite{dean-jensen,
  haskell-sedrakian,sedrakian-clark}. Neutron stars, which are
produced by a core-collapse supernova, are extremely dense objects,
made of highly degenerate asymmetric nuclear matter. While the
existence of a superfluid phase was theoretically predicted in the
early 1960s~\cite{migdal, ginzburg64} and a value of $\sim 1 \MeV$ was
correctly assigned to the pairing gap, a complete and systematic
theoretical description is yet an open problem. The existence of this
phase is significant from the point of view of
cooling~\cite{yakovlev-pethick, page-lattimer-prakash-steiner} and is
necessary to explain the observed glitches, which are sudden increases
in the rotational frequency of the star, followed by long relaxation
times to the pre-glitch values~\cite{anderson-itoh, pines-alpar,
  baym-pethick-pines-ruderman}. The attractive interaction is provided
by the nucleon-nucleon interaction and therefore, by analyzing the
two-body scattering phase shifts, one can conclude that neutrons can
pair in the singlet, $^1S_0$, channel at low densities, as they
prevail in the neutron-star crust~\cite{liv-rev}, and in the triplet,
$^3P_2-^3F_2$, channels ~\cite{maurizio-holt-finelli, sarath-ramanan,
  Drischler2016, Papakonstantinou2017} at higher densities as they are
expected in the neutron-star core. Protons pair up in the singlet
channel because their density never gets very high. However, proton
pairing is complicated by the interaction with the medium. In fact,
medium effects are very important for neutron pairing as well.

Even when modelled as pure neutron matter, a correct description of
the superfluid state is ridden with uncertainties. At low-densities,
in the singlet channel, the BCS approximation gives almost model
independent values for the transition temperature and gap, where by
the BCS approximation, we refer to solving the BCS gap equation using
free-space two-body interaction with a free single particle
spectrum. Hence, at the BCS level, phase-shift equivalence of the
two-body interaction suffices to yield model independent
results~\cite{Hebeler2007}. However, description of pairing in the
triplet channel, which typically occurs at higher densities becomes
challenging even at the BCS level, as the input free-space two-body
interactions are no-longer phase shift equivalent and hence the gaps
turn out to be model dependent~\cite{maurizio-holt-finelli,
  sarath-ramanan, Drischler2016, Papakonstantinou2017}. As already
noted, including the interaction of the neutrons with the medium is
very important and can lead to crucial medium modifications such as
screening of the free-space two-body interaction. However, the gap and
the transition temperature are extremely sensitive to the
approximation used to describe medium effects.

In our previous work~\cite{Ramanan2018}, the free-space
interaction was modified by including the effects of diagrams (a) and
(b) in Fig.~\ref{fig:diagram-RPA}.
\begin{figure}
\includegraphics[scale=0.9]{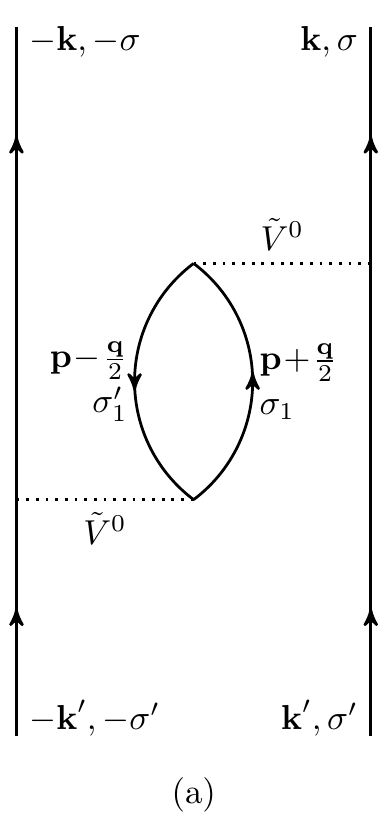}\hspace{1cm}
\includegraphics[scale=0.9]{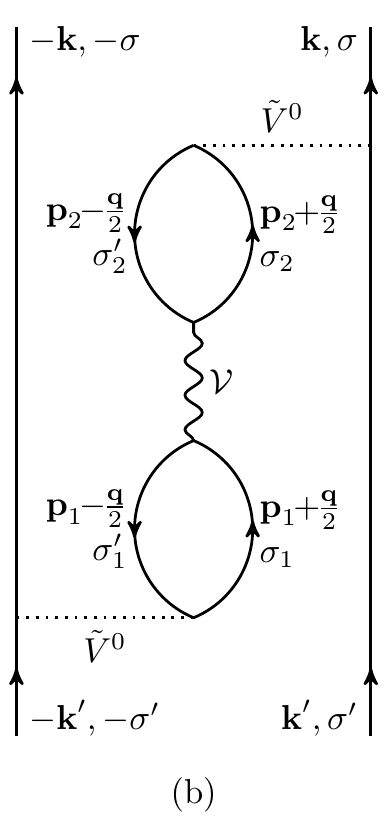}
\caption{\label{fig:diagrams}Feynman diagrams representing the induced
  interaction. The wiggly line in diagram (b) is meant to include the
  RPA bubble summation (see \Fig{fig:diagram-RPA}).}
\end{figure}
\begin{figure}
\includegraphics[scale=0.9]{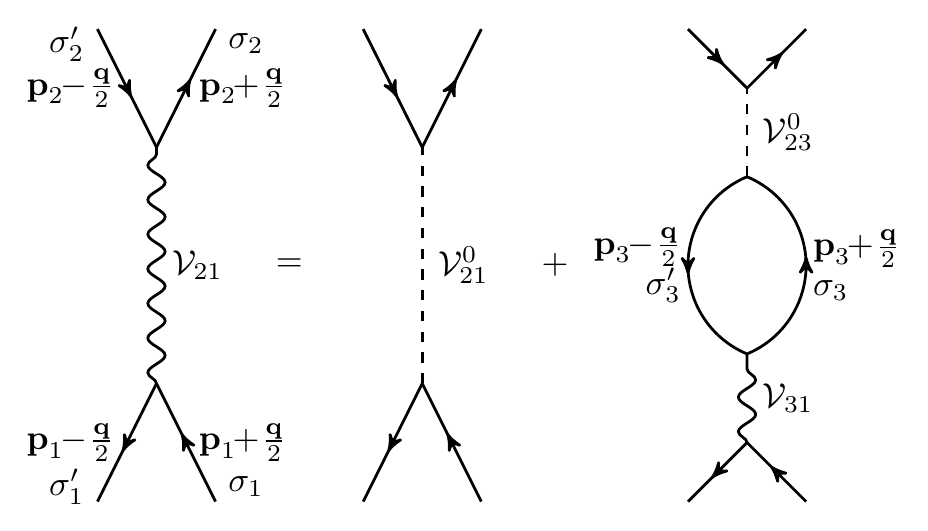}
\caption{\label{fig:diagram-RPA} Diagrammatic representation of the
  RPA, \Eq{eq:RPA}.}
\end{figure}
There, we used the free-space renormalized two-body interaction,
$\vlowk$, for the bare pairing interaction (that is, particle-particle
vertex, without medium correction) and for the 3p1h
(3-particle-1-hole) and 1p3h interactions, while the RPA (random-phase
approximation) series (see \Fig{fig:diagram-RPA}) used the
particle-hole (ph) interaction within the Landau approximation. The
Fermi-liquid parameters were obtained using phenomenological (Skyrme,
Gogny) energy-density functionals. The inclusion of diagram (b) is
important at higher densities since it reduces the effect of diagram
(a). In fact, it even results in the gap being anti-screened for $k_F
\gtrsim 0.7 \fmi$ ($k_F$ denotes the Fermi momentum which is related
to the density by $\rho = k_F^3/3\pi^2$ for pure neutron
matter). While the result that the effect of screening is reduced as
compared to diagram (a) is consistent with Quantum Monte Carlo (QMC)
calculations~\cite{Gandolfi2008,Abe2009,Gezerlis2010}, it is important
to check the dependence of the contribution from diagram (b) on the
approximation used. To our knowledge, all present calculations of
medium-polarization effects that include the RPA resummation,
e.g.,~\cite{Shen2005,Cao2006,Ding2016,Zhang2016}, rely on the Landau
approximation. As already noted in~\cite{Ramanan2018}, the Landau
approximation is valid for momentum transfers $q \ll k_F$, which is
clearly not the case in diagram (b) where $0\le q \lesssim 2 \,k_F$.

In this paper, we once again revisit the issue of medium polarization
in pure neutron matter.  We continue to use the $\vlowk$ interaction
for the bare pairing interaction and for the 3p1h and 1p3h
vertices. But now, we evaluate the RPA series in diagram (b) exactly,
i.e., beyond the Landau approximation. This is possible since we use
ph interactions of the Skyrme type. We repeat our calculations with
different Skyrme parameterizations (SLy4 and SLy5 from the Saclay-Lyon
family~\cite{Chabanat1998} and BSk19, BSk20, and BSk21 from the recent
Brussels-Montreal family~\cite{Goriely2010}) so that we can assess the
dependence of the result on the choice of the parameterization.

This paper is organized as follows. In \Sec{sect:form-diag}, we
revisit the set up of diagrams (a) and (b), followed by the
calculation of RPA diagrams using Skyrme interactions in
\Sec{sect:form-RPA}. We discuss in detail the choice of the
cutoff for the $\vlowk$ interaction and also
the parameters used for the Skyrme interaction in
\Sec{sect:interaction-choice}. In \Sec{sect:results}, we present the
main results of the paper. We study the dependence of the screened
interaction on the Skyrme parameterization as well as compare our
results to those obtained in~\cite{Ramanan2018}. We also revisit the
effect of preformed pairs on the density
  dependence of the transition temperature. Finally, in
\Sec{sect:concl}, we present our concluding remarks. Some details of
the calculations have been moved to appendices.

\section{Pairing with medium polarization}
\label{sect:form-diag}
In general, to obtain the $^1S_0$ pairing gap $\Delta$ or the critical
temperature $T_c$, one considers the gap equation
\begin{equation}
\Delta_k = -\frac{2}{\pi}\int_0^\infty \!dk'\, k^{\prime\,2}
  V(k,k')
  \frac{\Delta_{k'}\tanh\big(\frac{E_{k'}}{2T}\big)}{2E_{k'}}\,,
\label{eq:gap}
\end{equation}
where $V(k,k') = \langle k|V_{^1S_0}|k' \rangle$ denotes
the matrix element of the $nn$ interaction in the $^1S_0$ partial wave
for in- and outgoing momenta $k'$ and $k$, $E_k =
\sqrt{(\epsilon_k-\mu)^2+\Delta_k^2}$ is the
  quasiparticle energy with $\epsilon_k = k^2/2m^*$,
  $m^*$ is the neutron effective mass, $\mu$ is the effective chemical
  potential including the mean-field energy shift, and $T$ is the
  temperature. Including medium polarization effects in a way
  analogous to Debye screening of the Coulomb interaction
  \cite{FetterWalecka}, the interaction $V(k,k')$ can be written as
\begin{equation}
  V(k,k') = V^0(k,k')+V^{(a)}(k,k')+V^{(b)}(k,k')\,,
  \label{eq:V0ab}
\end{equation}
where $V^0$ is the contribution of the bare $nn$ interaction,
$V^{(a)}$ is the contribution of a single ph bubble exchange, and
$V^{(b)}$ represents the RPA resummation of the series of two and more
bubbles, see \Fig{fig:diagrams}.  To evaluate $V^{(a)}$, we proceed as
in \cite{Ramanan2018} and construct the bare interaction from
 \vlowk{}~\cite{Bogner2007}, given in
partial waves, as
\begin{multline}
\langle \vek{k}_1 \sigma_1, \vek{k}_2 \sigma_2|V^0
  |\vek{k}'_1 \sigma'_1, \vek{k}'_2 \sigma'_2\rangle =\\
    \sum_{s,m_s,m'_s} \sum_{l,l',m_l} \sum_{j}
      \CG{\half}{\sigma_1}{\half}{\sigma_2}{s}{m_s}
      \CG{\half}{\sigma'_1}{\half}{\sigma'_2}{s}{m'_s}
      \CG{l}{m_l}{s}{m_s}{j}{m_j}
      \CG{l'}{m'_l}{s}{m'_s}{j}{m_j}\\ \times 
      (4\pi)^2 i^{l'-l} Y^*_{lm_l}(\Omega_{\vek{Q}})Y_{l'm'_l}(\Omega_{\vek{Q}'})
      \langle Q|V^0_{sll'j}|
      Q'\rangle\,,
      \label{eq:partialwaves}
\end{multline}
with $m'_l = m_l+m_s-m'_s$, $m_j = m_l+m_s$, $\vek{Q} =
(\vek{k}_1-\vek{k}_2)/2$, and $\vek{Q}' =
(\vek{k}'_1-\vek{k}'_2)/2$. For the Clebsch-Gordan coefficients, we
follow the notation of the book by Varshalovich
\cite{Varshalovich}. The antisymmetrized interaction is defined by
\begin{equation}
  \langle 1,2|\tilde{V}^0|1',2'\rangle = 
  \langle 1,2|V^0|1',2'\rangle - \langle 1,2|V^0|2',1'\rangle
\end{equation}
where the short-hand notation $1$ stands for $\kv_1\sigma_1$ etc.

The expression corresponding to diagram (a) reads
\begin{multline}
  V^{(a)}(k,k') =
  \frac{-1}{4\pi}\sum_{\sigma\sigma'}
  \CG{\half}{\sigma}{\half}{-\sigma}{0}{0}
  \CG{\half}{\sigma'}{\half}{-\sigma'}{0}{0}
  \int\! \frac{d\Omega_{k}}{4\pi}
  \int\! \frac{d\Omega_{k'}}{4\pi}\\
  \times\sum_{\pv\sigma_1}
  \langle -\kv\, -\sigma, \pv+\tfrac{\qv}{2}\, \sigma_1|\tilde{V}^0
    |-\kv'\, -\sigma', \pv-\tfrac{\qv}{2}\, \sigma'_1\rangle\\
  \times \frac{n_{\pv-\frac{\qv}{2}}-n_{\pv+\frac{\qv}{2}}}{\epsilon_{\pv+\frac{\qv}{2}}-\epsilon_{\pv-\frac{\qv}{2}}}
  \langle \pv-\tfrac{\qv}{2}\, \sigma'_1, \kv \, \sigma|\tilde{V}^0
    |\pv+\tfrac{\qv}{2}\, \sigma_1, \kv'\, \sigma'\rangle\,.
  \label{eq:diagram-a}
\end{multline}
where $\qv = \kv-\kv'$ and $\sum_{\pv}$ stands for $\int
d^3p/(2\pi)^3$. The occupation numbers $n_{\pv}$ can be safely
approximated by step functions, $n_{\pv} = \theta(k_F-p)$, 
because the temperature $T$ and the pairing
gap $\Delta$ are much smaller than the Fermi energy $\epsilon_F$.

Similarly, the expression for diagram (b) can be written as
\begin{multline}
  V^{(b)}(k,k') =
  \frac{1}{4\pi}\sum_{\sigma\sigma'}
  \CG{\half}{\sigma}{\half}{-\sigma}{0}{0}
  \CG{\half}{\sigma'}{\half}{-\sigma'}{0}{0}
  \int\! \frac{d\Omega_q}{4\pi}
  \int\! \frac{d\Omega_{q'}}{4\pi}\\
  \times \sum_{\pv_1\sigma_1}\sum_{\pv_2\sigma_2}
  \langle -\kv \, -\sigma, \pv_1+\tfrac{\qv}{2} \, \sigma_1|\tilde{V}^0
    |-\kv' \, -\sigma', \pv_1-\tfrac{\qv}{2}\, \sigma'_1\rangle\\
  \times \frac{n_{\pv_1-\frac{\qv}{2}}-n_{\pv_1+\frac{\qv}{2}}}{\epsilon_{\pv_1+\frac{\qv}{2}}-\epsilon_{\pv_1-\frac{\qv}{2}}}
  \calV_{\pv_2-\frac{\qv}{2}\sigma'_2,\pv_2+\frac{\qv}{2}\sigma_2;\pv_1+\frac{\qv}{2}\sigma_1,\pv_1-\frac{\qv}{2} \sigma'_1}\\
  \times\frac{n_{\pv_2-\frac{\qv}{2}}-n_{\pv_2+\frac{\qv}{2}}}{\epsilon_{\pv_2+\frac{\qv}{2}}-\epsilon_{\pv_2-\frac{\qv}{2}}}
  \langle \pv_2-\tfrac{\qv}{2} \, \sigma'_2, \kv \, \sigma|\tilde{V}^0
    |\pv_2+\tfrac{\qv}{2}\, \sigma_2, \kv' \, \sigma'\rangle\,.
  \label{eq:diagram-b}
\end{multline}
Here, $\calV$ denotes the ph interaction including the resummation of
bubble diagrams. It is different from the particle-particle (pp)
interaction $V^0$ and we therefore use a different notation. Like the
effective mass $m^*$, the ph interaction will not be derived from the
free-space interaction $V^0$, but from a phenomenological Skyrme
energy functional. Previous studies of screening used the simplest
approximation to $\calV$, namely the lowest-order Landau approximation
\begin{multline}
\calV_{\pv_2-\frac{\qv}{2}\sigma'_2,\pv_2+\frac{\qv}{2}\sigma_2;\pv_1+\frac{\qv}{2}\sigma_1,\pv_1-\frac{\qv}{2} \sigma'_1}
\approx
\frac{f_0\, \delta_{\sigma'_1\sigma_1}\delta_{\sigma_2\sigma'_2}}{1-f_0\Pi_0(q)}\\
+\frac{g_0\, \sigv_{\sigma'_1\sigma_1}\dotp\sigv_{\sigma_2\sigma'_2}}{1-g_0\Pi_0(q)}\,,
\label{eq:Landauapproximation}
\end{multline}
where $f_0$ and $g_0$ are the $L=0$ Landau parameters in the density
and spin channel, respectively, $\sigv$ are the Pauli matrices, and
$\Pi_0(q)$ is the Lindhard function in the static limit, see
Appendix~\ref{app:Lindhard}. Inserting \Eq{eq:Landauapproximation}
into \Eq{eq:diagram-b} and renaming $\kv,\kv',\qv,\pv_i\to
\qv,\qv',\kv,\pv_i-\kv/2$, one recovers the expression given in our
previous work \cite{Ramanan2018}. The aim of the present study is to
go beyond this approximation and to include the RPA calculated with
the full Skyrme ph interaction.

\section{RPA with Skyrme interaction}
\label{sect:form-RPA}
The RPA with Skyrme interactions has been extensively studied, see the
review \cite{Pastore2015}. However, in most calculations, one is
interested in the response function and not in the vertex $\calV$
which we need here. The RPA vertex function for Skyrme-like
interactions was, e.g., considered in \cite{Garcia1992}.

The residual ph interaction is derived from the Skyrme energy
functional $E$ as \cite{Garcia1992}
\begin{multline}
  \calV^0_{\pv_2-\frac{\qv}{2}\,\sigma_2',\pv_2+\frac{\qv}{2}\,\sigma_2;
       \pv_1+\frac{\qv}{2}\,\sigma_1,\pv_1-\frac{\qv}{2}\,\sigma_1'}\\
  = \frac{\delta^2 E}
    {\delta\rho_{\pv_2-\frac{\qv}{2}\,\sigma_2',\pv_2+\frac{\qv}{2}\,\sigma_2}
      \delta\rho_{\pv_1+\frac{\qv}{2}\,\sigma_1,\pv_1-\frac{\qv}{2}\,\sigma_1'}}\,,
    \label{eq:functionalderivative}
\end{multline}
where $\rho_{{\pv}\,\sigma,\pv',\sigma'}$ denotes the density matrix
with the momentum and spin indices as defined in
\Fig{fig:diagram-RPA}. Let us introduce the short-hand notation
\begin{gather}
\calV^0_{21} = \calV^0_{\pv_2-\frac{\qv}{2}\,\sigma'_2,\pv_2+\frac{\qv}{2}\,\sigma_2;
    \pv_1+\frac{\qv}{2}\,\sigma_1,\pv_1-\frac{\qv}{2}\,\sigma'_1}\,,\\
\sigv_1 = \sigv_{\sigma_1'\sigma_1}\,,
\end{gather}
etc., where $1$ stands for the quantum numbers
$\pv_1,\sigma_1,\sigma'_1$ and similarly for $2$. If no spin operator
$\sigv_1$ is written, the corresponding term is assumed to be
proportional to $\delta_{\sigma_1'\sigma_1}$. Using this notation, the
interaction derived from a standard Skyrme functional has the form
\begin{multline}
  \calV^0_{21} = v^0_{1}(q) + v^0_{2}\, (\pv_1-\pv_2)^2\\
  + [v^0_{4}(q)+v^0_{5}\,(\pv_1-\pv_2)^2]\,\sigv_1\dotp\sigv_2\\
  + v^0_{8}\,i\qv\dotp (\pv_1-\pv_2)\crossp(\sigv_1+\sigv_2)\,.
  \label{eq:Skyrmeforce}
\end{multline}
The $v^0_{i}$ can be density dependent and their expressions in terms of
the parameters of the Skyrme force are given in
Appendix~\ref{app:Skyrme}. Notice that, as a consequence of its
derivation via \Eq{eq:functionalderivative}, $\calV^0_{21}$ contains
already both the direct and the exchange term.

The RPA vertex $\calV$ is obtained from the ph
interaction $\calV^0$ by solving the Bethe-Salpeter like equation (see
\Fig{fig:diagram-RPA})
\begin{equation}
  \calV_{21} = \calV^0_{21}-\sum_3 \calV^0_{23}G^0_{\ph}(\pv_3,\qv)\calV_{31}\,,
  \label{eq:RPA}
\end{equation}
where the minus sign comes from the closed fermion loop, $\sum_3$
is a short-hand notation for $\sum_{\sigma_3\sigma'_3}\int d^3p_3/(2\pi)^3$, and
\begin{equation}
  G^0_{\ph}(\pv,\qv) = \frac{n_{\pv-\frac{\qv}{2}}-n_{\pv+\frac{\qv}{2}}}
  {\epsilon_{\pv+\frac{\qv}{2}}-\epsilon_{\pv-\frac{\qv}{2}}}
\end{equation}
is the ph propagator in the static limit, i.e., for $\omega\to 0$,
where $\omega$ is the total energy of the ph pair. While \Eq{eq:RPA} is in general quite difficult to solve,
it is very simple in the case of a Skyrme interaction.

The RPA vertex has a more general structure than the Skyrme ph
interaction in \Eq{eq:Skyrmeforce}. Nevertheless, since all terms in
the Skyrme force are at most quadratic in $\pv_1$ and $\pv_2$, only a
finite number of terms are generated \cite{Garcia1992}. The number of
independent terms is further reduced because the vertex is Hermitian,
i.e., $\calV_{21} = \calV^*_{12}$, and the dependence on the angles of
$\pv_1$ and $\pv_2$ is at most $L=1$. In the case of a standard Skyrme
interaction, it turns out that the following ansatz for $\calV$ is
sufficient:
\begin{multline}
  \calV_{21} = v_{1}
  + v_{2}\, (p_1^2 + p_2^2)
  + v_{3}\, \pv_1\dotp\pv_2
  + v_{4}\,\sigv_1\dotp\sigv_2\\
  + v_{5}\,\sigv_1\dotp\sigv_2\, (p_1^2 + p_2^2)
  + v_{6}\,\sigv_1\dotp\sigv_2\, \pv_1\dotp\pv_2
  + v_{7}\,\sigv_1\dotp\qv\, \sigv_2\dotp\qv\\
  + v_{8}\,i\qv\dotp (\pv_1\crossp\sigv_1-\pv_2\crossp\sigv_2)
  + v_{9}\,i\qv\dotp (\pv_1\crossp\sigv_2-\pv_2\crossp\sigv_1)\\
  + v_{10}\,p_1^2 p_2^2
  + v_{11}\,\pv_1\dotp\qv\,\pv_2\dotp\qv
  + v_{12}\,\sigv_1\dotp\sigv_2\, p_1^2 p_2^2\\
  + v_{13}\,\sigv_1\dotp\sigv_2\,\pv_1\dotp\qv\,\pv_2\dotp\qv
  + v_{14}\,\sigv_1\dotp\qv\, \sigv_2\dotp\qv\, (p_1^2+p_2^2)\\
  + v_{15}\,\sigv_1\dotp\qv\, \sigv_2\dotp\qv\, p_1^2 p_2^2
  + v_{16}\,i\qv\dotp(\pv_1\crossp\sigv_1\, p_2^2-\pv_2\crossp\sigv_2\, p_1^2)\\
  + v_{17}\,i\qv\dotp(\pv_1\crossp\sigv_2\, p_2^2-\pv_2\crossp\sigv_1\, p_1^2)\\
  + v_{18}\,\qv\dotp\pv_1\crossp\sigv_1\,\qv\dotp\pv_2\crossp\sigv_2\,.
\label{eq:v1-18}
\end{multline}
The coefficients $v_{i}$ are functions of $q$ and $\rho$, but we drop
the arguments for brevity. In order to determine the functions $v_{i}$,
we insert \Eqs{eq:Skyrmeforce} and~(\ref{eq:v1-18}) into
\Eq{eq:RPA}. The integrals over $\pv_3$ can be expressed in terms of
the generalized Lindhard functions
\begin{gather}
  \Pi_n(q) = -2\int \frac{d^3p}{(2\pi)^3} p^n G_{\ph}\,,\\
  \Pi_{2L}(q) = -2\int \frac{d^3p}{(2\pi)^3} p^2\cos^2\theta\, G_{\ph}\,,\\
  \Pi_{2T}(q) = -\int \frac{d^3p}{(2\pi)^3} p^2\sin^2\theta\, G_{\ph}
     = \frac{\Pi_2-\Pi_{2L}}{2}\,,\label{eq:Pi2T}
\end{gather}
$\theta$ being the angle between $\pv$ and $\qv$. The explicit
expressions for these functions are given in
Appendix~\ref{app:Lindhard}. Collecting the coefficients of the
different operators that appear in \Eq{eq:v1-18}, one obtains a linear
system of equations for the $v_{i}$ of the form
\begin{equation}
  v_{i} = v^0_{i} + \sum_j A_{i,j} v_{j}\,,
  \label{eq:RPAmatrixform}
\end{equation}
where the matrix elements $A_{i,j}$ are products of the different
$v^0_{i}$ and $\Pi_i$. Solving this system of equations, one obtains
analytical expressions for the $v_{i}$, which are listed in
Appendix~\ref{app:RPA}.

To get a qualitative idea about the difference between the full RPA
and the Landau approximation, let us consider the static ($\omega\to
0$) density response (ph spin $S=0$)
\begin{multline}
  \Pi_{\RPA}^{(0)} =
    -\sum_1 G^0_{\ph}(\pv_1,\qv) \\+ \sum_{1,2} G^0_{\ph}(\pv_1,\qv)
    V_{12} G^0_{\ph}(\pv_2,\qv)\\
  = \Pi_0 + v_{1}\, \Pi_0^2 + 2 v_{2}\, \Pi_0 \Pi_2 + v_{10}\, \Pi_2^2\,.
\end{multline}
Similarly, one obtains the spin response ($S=1$) by including Pauli
matrices into the sums over 1 and 2. The result for the transverse
spin response ($M=\pm 1$, where $M$ denotes the projection of the ph
spin on the direction of $\qv$) takes the form
\begin{equation}
  \Pi_{\RPA}^{(1,\pm 1)}
  = \Pi_0 + v_{4}\, \Pi_0^2 + 2 v_{5}\, \Pi_0 \Pi_2 + v_{12}\, \Pi_2^2\,,
\end{equation}
whereas additional terms appear in the case of the longitudinal spin
response ($S=1$, $M=0$):
\begin{equation}
  \Pi_{\RPA}^{(1,0)} = \Pi_{\RPA}^{(1,\pm 1)}
  + q^2(v_{7}\,\Pi_0^2 + 2 v_{14}\, \Pi_0 \Pi_2 + v_{15}\, \Pi_2^2)\,.
\end{equation}
In the absence of tensor terms, which we have not considered here
because they do not appear in standard Skyrme interactions, these
expressions agree with Eqs.~(53), (55), and (56) of
\Ref{Pastore2015}. For comparison, in Landau approximation, one has
$v_{1} = f_0/(1-f_0\Pi_0)$, $v_{4} = g_0/(1-g_0\Pi_0)$, and all other
$v_{i}=0$, and therefore
\begin{equation}
  \Pi_{\text{Landau}}^{(S=0)} = \frac{\Pi_0}{1-f_0\Pi_0}\,,\quad
  \Pi_{\text{Landau}}^{(S=1)} = \frac{\Pi_0}{1-g_0\Pi_0}\,.
\end{equation}

\begin{figure}
  \includegraphics[width=0.99\columnwidth]{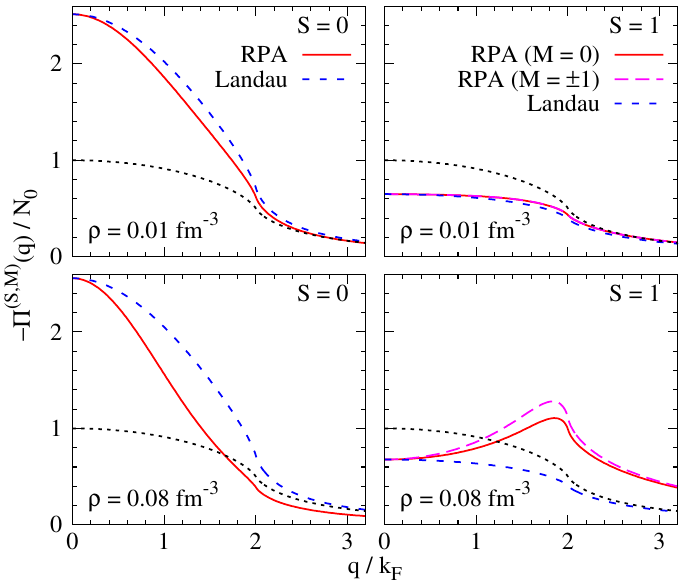}
\caption{\label{fig:RPA-Landau-compare} Density response ($S=0$, left
  column) and longitudinal ($M=0$) and transverse ($M=\pm 1$) spin
  responses ($S=1$, right column) in units of the density of states
  $N_0 = m^* k_F/\pi^2$ as functions of $q/k_F$, for two different
  densities: $\rho = 0.01\fm^{-3}$ ($k_F = 0.67\fmi$, upper row)
  and $0.08\fm^{-3}$ ($k_F = 1.33\fmi$, lower row), obtained with
  the SLy5 interaction. Results of the full RPA (red solid line and
  long purple dashes) are compared with those of the Landau
  approximation (short blue dashes) and with the Lindhard function
  $-\Pi_0$ (black dotted line).}
\end{figure}

The full RPA responses and the responses in Landau
approximation are shown in \Fig{fig:RPA-Landau-compare}
for two different densities. At low density ($\rho = 0.01\fm^{-3}$,
upper panels), the Landau approximation reproduces quite well the full
RPA result. This is because the integration region of the internal
loop momenta $p$ as well as the relevant range of external momenta $q$
scale with $k_F$, so that the momentum dependent terms of the full
Skyrme interaction $\calV^0$ are small at low density. Nevertheless,
one can see that with increasing momentum the RPA responses are closer
to the Hartree-Fock response (Lindhard function $\Pi_0$) than the
Landau approximation. This reflects the fact that, roughly speaking,
$\calV^0$ becomes smaller with increasing momentum, as one would
expect for a finite-range interaction which is simulated by the
momentum dependence of the Skyrme force. At higher density
($\rho=0.08\fm^{-3}$, lower panels of \Fig{fig:RPA-Landau-compare}), the
momentum dependence of the interaction is so strong that it even
changes sign. As a consequence, the density response (left panel)
falls below $\Pi_0$ at large $q/k_F$, while the spin response (right
panel) becomes enhanced compared to $\Pi_0$. It is not clear whether
this is realistic or just an artefact of the limitation of the Skyrme
force to terms up to second order in the momenta.

It would be interesting to compare these results with those obtained
with a true finite-range interaction such as the Gogny force. However,
in that case the solution of the RPA is very difficult and this is
beyond the scope of the present study. QMC
calculations~\cite{Buraczynski2016} seem to indicate that the static
density response of neutron matter should approach the free one (i.e.,
$\Pi_0$ but computed without effective mass; note that in general
$m^*$ is momentum dependent) at large $q$.
\section{Choice of interactions and parameters}
\label{sect:interaction-choice}
Before presenting numerical results, let us specify the choice of the
interaction $V^0$ in the pp channel, which is also used in the
3p1h and 1p3h vertices in diagrams (a) and (b), and of the
parameterizations of the Skyrme interaction $\calV^0$ used in the
calculation of the effective mass $m^*$ and in the ph channel for the
RPA.

As bare interaction $V^0$, we use the low-momentum interaction
\vlowk{} derived from the AV$_{18}$ interaction by a
renormalization-group evolution~\cite{Bogner2007}. The partial waves
[cf. \Eq{eq:partialwaves}] are summed up to $j_{\text{max}}=3$ which
for our purposes is sufficient to reach converged results
\cite{Ramanan2018}.

The \vlowk{} interaction has an additional parameter, namely the
momentum cutoff $\Lambda$. Although, by construction, \vlowk{} gives
cutoff-independent results in the two-body sector (at sufficiently low
energies), this is not the case if it is used in a many-body
calculation, where the dependence on the cutoff indicates missing
medium and higher-body corrections. Actually, the purpose of using
\vlowk{} instead of AV$_{18}$ is to make the interaction ``more
perturbative'' and thus more suitable for approximations used in the
many-body problem. A common choice is a cutoff $\Lambda = 2\fmi$ which
we will also use here, especially at densities with $k_F\gtrsim
0.8\fmi$.

However, as explained in detail in \Ref{Ramanan2018}, to reproduce the
correct low-density limit of the screening correction
(Gor'kov-Melik-Barkhudarov (GMB) result~\cite{Gorkov1961}), it is
important that the 3p1h and 1p3h vertices approach the $nn$ scattering
length, while in diagrams (a) and (b) these vertices contain the
interaction only to leading order (Born term). We remind the reader
that in the limit of small cutoffs $\Lambda$ and small momenta $k$ and
$k'$, the interaction $V^0(k,k')$, the cutoff $\Lambda$,
and the scattering length $a$ are roughly related by
\begin{equation}
  V^0\sim \Big(\frac{m}{a}-\frac{2m\Lambda}{\pi}\Big)^{-1}\,.
\end{equation}
Hence, by lowering the cutoff as much as possible, we can achieve
$V^0\sim a/m$. Obviously, the cutoff must remain larger than $k_F$ if
one wants to describe Cooper pairing. In practice, when solving the
gap equation with $V^0$ alone, one can decrease the cutoff to $\Lambda
= 2.5\, k_F$ (with a regulator of the form $e^{-(k^2/\Lambda^2)^5}$)
without affecting the critical temperature. As in \Ref{Ramanan2018},
we will use this as an alternative choice, especially for low
densities with $k_F\lesssim 0.8\fmi$.

The reason why we do not use the density dependent cutoff at higher 
values of $k_F$ is the following.
As $k_F$ increases, the variable cutoff grows
and as a result will include the coupling between the low- and
high-momentum physics, making the interaction less useful in
perturbation theory. The BCS transition temperatures and 
gaps in the singlet channel are unaffected by the coupling
between low- and high-momentum modes, as they depend on the correct
reproduction of the two-body scattering data. But the perturbativeness
of the interaction becomes important in the 3p1h and 1p3h
  vertices, and also in the Nozi\`eres-Schmitt-Rink (NSR)
approach (see next section and Appendix~\ref{app:NSR}), where
one has to compensate for the double counting of the
  single-particle energy shift (see~\cite{Ramanan2013}), which
in our case is done within the Hartree-Fock approximation.
Therefore, it is crucial to soften the
interaction. Hence we use a variable cutoff of $2.5\,k_F$ until
$k_F \approx 0.8 \fmi$ and a constant
cutoff of $2 \fmi$ for higher densities. 

Concerning the ph interaction, there are hundreds of different
parameterizations of the Skyrme interaction on the market, which were
fitted in different ways and for different purposes. Fortunately, the
number of interactions suitable for neutron matter is much
smaller. The first interactions that were fitted to reproduce not
only finite nuclei but also infinite neutron matter were those of the
Saclay-Lyon (SLy\dots) family. We will use the parameterizations SLy4
and SLy5 \cite{Chabanat1998}. More recent Skyrme interactions
developed specifically for astrophysical applications are those of the
Brussels-Montreal (BSk\dots) family. Out of this family, we will use
the parameterizations BSk19, BSk20, and BSk21 \cite{Goriely2010}. To
use several different parameterizations can be useful to get an idea
how strongly the results depend on this choice.

The Fermi-liquid parameters corresponding to these interactions (see
Appendix~\ref{app:FL}) are shown in \Fig{fig:FL}.
\begin{figure}
\includegraphics[width=0.99\columnwidth]{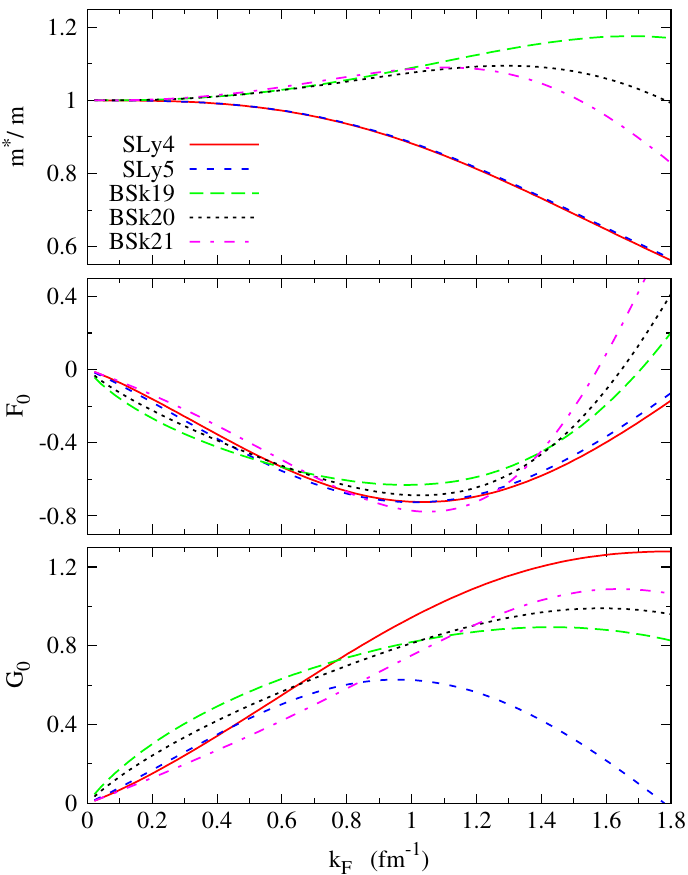}
\caption{\label{fig:FL} Fermi-liquid parameters $m^*$, $F_0$, and
  $G_0$ as functions of the Fermi momentum $k_F$ for five different
  Skyrme parameterizations SLy4, SLy5, BSk19, BSk20, and BSk21.}
\end{figure}
In our preceding work~\cite{Ramanan2018}, we computed the Fermi-liquid
parameters from the SLy4 parameterization, but using $\eta_J = 1$ in the calculation of $G_0$,
although the SLy4 functional was fitted without the $\JJ^2$ terms,
i.e., with $\eta_J=0$ (following the notations of
  \cite{Bender2003}). This was somewhat inconsistent (although quite
common in the literature). Here, by using the SLy5 parameterization,
which was fitted with $\eta_J=1$, we avoid this inconsistency and we
still find Fermi liquid parameters (blue short dashes) very close to
those shown in our preceding work (Fig. 3 of \Ref{Ramanan2018}). On
the contrary, the SLy4 results shown here (red solid lines) are now
obtained with $\eta_J=0$ which explains why the $G_0$ Landau parameter
is much more repulsive than with SLy5.

In \Ref{Chamel2010}, it was pointed out that the $G_0$ parameter of
SLy5 is probably too small, which can lead to ferromagnetic
instabilities at higher densities. It was therefore suggested that one
should rather use SLy4 (with $\eta_J=0$) or more recent
parameterizations such as BSk19-21 which were fitted (again with
$\eta_J=0$) to give a reasonable $G_0$ at saturation density. But even
the latter parameterizations lead to instabilities in the spin channel
for neutron matter above saturation density $0.16\fm^{-3}$
\cite{Pastore2014}.

\section{Numerical results}
\label{sect:results}
Let us now proceed as in \Ref{Ramanan2018} to compute the effect of
the exchange of RPA excitations on pairing. We first compute the
modified pairing interaction [\Eqs{eq:V0ab}, (\ref{eq:diagram-a}), and
(\ref{eq:diagram-b})]. The summations and integrations in \Eqs{eq:diagram-a} and
(\ref{eq:diagram-b}) are all done numerically. The resulting matrix
elements $V(k,k')$ are then used in the gap equation
(\ref{eq:gap}) to compute the critical temperature or the gap. Here,
we will discuss $T_c$, from which the gap can be obtained to very good
precision from $\Delta_{k_F}(T=0) \approx 1.76\, T_c$
\cite{FetterWalecka}.

In our preceding work \cite{Ramanan2018}, using the Landau
approximation, we found that there are strong cancellations between
the attractive exchange of density fluctuations ($S=0$) and the
repulsive exchange of spin fluctuations ($S=1$). While at low density,
the repulsive effect was dominant, it turned out that at higher
densities, the $S=0$ contribution became dominant due to its
enhancement by the negative Landau parameter $f_0$ and the suppression
of the $S=1$ contribution by the positive Landau parameter $g_0$.

\begin{figure}
\includegraphics[width=0.99\columnwidth]{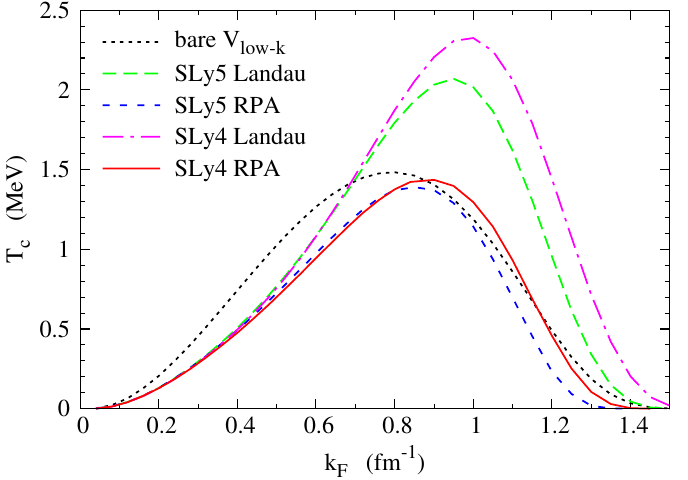}
\caption{\label{fig:Tc-Lan-RPA} Critical temperature $T_c$ as a
  function of $k_F$, obtained with the bare \vlowk{} (cutoff
  $\Lambda=\min(2.5k_F,2\fmi)$, $m^*$ computed with SLy5) and
  with medium polarization calculated either within the Landau
  approximation or within the full RPA, with two different
  parameterizations of the Skyrme interaction (SLy4 and SLy5).}
\end{figure}

Let us see how this picture is modified when one includes the full
Skyrme RPA instead of the Landau approximation. In \Fig{fig:Tc-Lan-RPA},
we display a comparison of the critical temperatures obtained with
different levels of approximation. The black dotted curve represents
the result without medium polarization. The long green dashes include
the medium polarization computed within the Landau approximation using
the SLy5 interaction. One clearly sees the suppression of $T_c$ at low
density due to screening, gradually turning into an enhancement due to
anti-screening at $k_F\gtrsim 0.7\fmi$. This curve is very similar
to the result we obtained in our previous work~\cite{Ramanan2018},
while the one obtained with SLy4 (purple dashed-dotted curve) shows
even stronger anti-screening at high density because of the larger
value of the $G_0$ Landau parameter (see discussion in
\Sec{sect:interaction-choice}). The short blue dashed (SLy5) and the
red solid (SLy4) lines are the corresponding results obtained within
the full RPA instead of the Landau approximation. We see that for
$k_F\lesssim 0.4\fmi$, the full RPA and the Landau approximation
agree very well. But at higher densities, the critical temperature
within full RPA is always lower than within the Landau
approximation. In the case of SLy5, the full RPA never gives
anti-screening (i.e., enhancement of $T_c$ compared to the bare \vlowk{}
interaction) and in the case of SLy4, anti-screening survives only in
some range of densities around $k_F\sim 1\fmi$ and it is much
weaker than within the Landau approximation.

Qualitatively, this result can be understood by looking at
\Fig{fig:RPA-Landau-compare}. One sees two effects acting in the same
direction: First, within the full RPA, the density response is less
enhanced than within the Landau approximation, and, second, the spin
response is less suppressed or even enhanced compared to the free
one. (Strictly speaking, because of the spin-orbit interaction, the
spin of the ph excitation is not a good quantum number any more, but
nevertheless the argument remains qualitatively valid.)

\begin{figure}
\includegraphics[width=0.99\columnwidth]{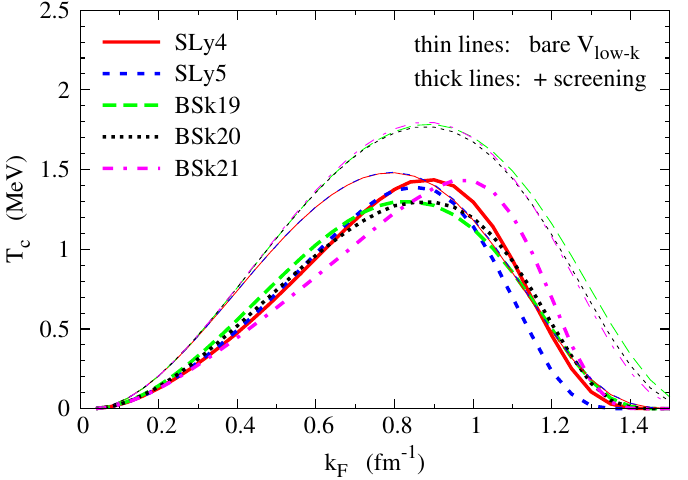}
\caption{\label{fig:Tc-SLy-BSk} Thin lines: critical temperature $T_c$
  as a function of the Fermi momentum $k_F$, obtained with the the
  bare \vlowk{} [cutoff $\Lambda=\min(2.5k_F,2\fmi)$] using the
  effective mass $m^*$ of the same Skyrme parameterizations as in
  \Fig{fig:FL}. Thick lines: corresponding results including the
  medium polarization computed within full RPA.}
\end{figure}

The question arises whether this is a specific property of the SLy
interactions or whether similar results are obtained with other Skyrme
parameterizations. In \Fig{fig:Tc-SLy-BSk},
we compare the critical temperatures obtained with the bare \vlowk{}
(thin lines) and with medium polarization within full RPA (thick
lines) for five different Skyrme parameterizations (SLy4, SLy5, BSk19,
BSk20, and BSk21). Looking at the thin lines, one sees that already at
the level of the bare \vlowk{}, the SLy and BSk parameterizations give
quite different density dependence of $T_c$. This can easily be
understood from \Fig{fig:FL}: the SLy parameterizations predict a
lower effective mass $m^*$ in neutron matter than the BSk ones. Since
the gap and $T_c$ depend exponentially on the density of states
$N_0=m^* k_F/\pi^2$, this has a dramatic effect, especially in the
weak-coupling regime, i.e., at high density. Concerning the results
with medium polarization (thick lines), none of the BSk forces gives
anti-screening. At high densities, screening is strongest with BSk19
and weakest with BSk21, while at low densities, it is the
opposite. There is a clear relationship between screening and the
$G_0$ parameter: the more repulsive $G_0$ is, the weaker is the
screening. This general trend is easily understood within the Landau
approximation but apparently it also survives in the full RPA. A quite
surprising result is that, at least if one compares these two families
of Skyrme parameterizations, the model dependence with screening is
weaker than without screening. Of course, this may be accidental.

In addition to the screening by the medium, there are correlations
between neutrons above $T_{c}$ and such correlations can be taken into
account via the Nozi\`eres-Schmitt-Rink (NSR)
approach~\cite{Nozieres1985}, wherein, for a given chemical potential
$\mu$, the density of the interacting neutrons is enhanced by the
correlations calculated within the ladder approximation. As a result,
the total density, $\rho_{\tot}$ is written as a sum of uncorrelated
and correlated densities,
\begin{equation}
 \rho_{\tot} = \rho_0 + \rho_{\corr}\,.
 \label{eq:rhotot}
\end{equation}
The formulas for $\rho_0$ and $\rho_{\corr}$ are given in
  Appendix~\ref{app:NSR}. This effect is important at
low densities where $T_c/\mu$ is not too small (strong coupling regime), 
as already seen in~\cite{Ramanan2013,Ramanan2018}. Since $T_c$ is computed 
according to the gap equation~(\ref{eq:gap}) as a function of $\mu$, but the
relation between $\mu$ and $\rho_\tot$ is changed, this implies that
the dependence of $T_c$ on $\rho_\tot$ (or on $k_{F,\tot} =
(3\pi^2\rho_{\tot})^{1/3}$) is changed, too. Notice that now, for
given $\rho_{\tot}$ (or $k_{F,\tot}$) the relation $\Delta(T=0) =
1.76 \,T_c$, is no longer true.

In~\cite{Ramanan2018}, we already studied the combined effect of
correlations using the NSR approach, and screening
in the Landau approximation. Since we have seen that the screening
changes completely if one passes from the Landau approximation to the
full Skyrme RPA, we would like to revisit this study using the Skyrme
RPA instead of the Landau approximation.

\begin{figure}[t]
  \includegraphics[width = 0.99\columnwidth]
                  {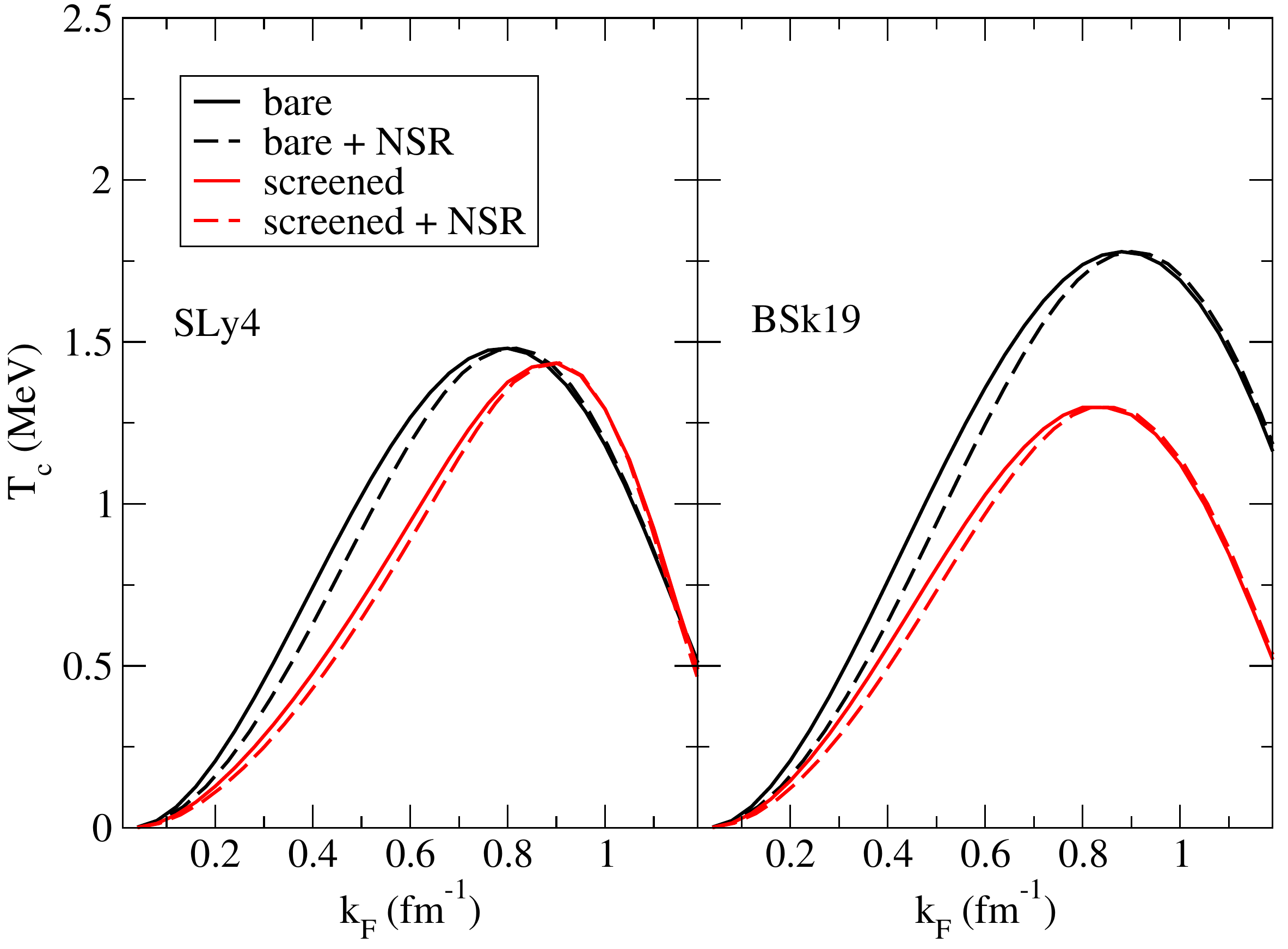}     
  \caption{Transition temperature versus $k_F$ or $k_{F,\tot}$,
    respectively, obtained with (dashed lines) and without (solid
    lines) pair correlations via the Nozi\`eres-Schmitt-Rink
    procedure, both with the bare (black lines) and with the screened
    (red lines) interactions. All curves were obtained including the
    effective mass $m^*$ computed with the respective Skyrme
    interaction (SLy4 in the left panel and BSk19 in the right
    panel).}
   \label{fig:NSR}
\end{figure}

Figure~\ref{fig:NSR} shows the transition temperatures including the
effect of preformed pairs via the NSR approach (i.e., as functions of
$k_{F,\tot}$), obtained with the bare and the screened interactions,
respectively (black and red dashed lines). For ease of comparison, the
figure also displays the transition temperatures calculated from the
bare and the screened interaction without the NSR effect (i.e., as
functions of $k_F$; black and red solid lines). All curves take
into account the effective mass, which has been calculated using the
respective Skyrme parameterizations (SLy4 in the left panel and BSk19
in the right panel).

As already observed in \cite{Ramanan2018}, we note that the effect of
screening is much stronger than the effect of pair correlations,
except at very low densities ($k_F\lesssim 0.1 \fmi$ in the case of
SLy4 and $k_F\lesssim 0.2 \fmi$ in the case of BSk19), where both
effects are equally important (cf. black dashed and red solid lines).

As expected, the effect of preformed pairs is in fact limited to the
range $k_F\lesssim 0.8\fmi$, where the ratio $T_c/\mu$ is not too
small so that the neutron matter is close to the so-called BCS-BEC
crossover regime \cite{Matsuo2006,Margueron2007}. This regime has been
extensively studied with ultracold atoms \cite{Strinati2018} and in
that case it was shown in \Ref{Pisani2018} that by including
simultaneously the effects of screening and pairing fluctuations
(corresponding to the non-condensed preformed pairs), one reproduces
very well the experimental result for the critical temperature in the
unitary limit, i.e., in the case of a contact interaction with
infinite scattering length.

In the present case of neutron matter, if the effect of pair
correlations is included on top of screening (red dashed lines in
Fig.~\ref{fig:NSR}), it is even weaker than in the case without
screening, i.e., the difference between the solid and dashed red
curves is smaller than the difference between the solid and dashed
black curves, which do not include screening. This is because the
screening weakens the attractive interaction, reducing the pairing
correlations and hence the correlated density compared to the one
obtained with the bare interaction, in agreement with our conclusions
in our previous work (see Fig.~15 in~\cite{Ramanan2018}).

\section{Conclusion}
\label{sect:concl}
The main goal of this work is to check the Landau approximation used
in our previous work~\cite{Ramanan2018} (and in other recent studies
of medium polarization effects
\cite{Shen2005,Cao2006,Ding2016,Zhang2016}) for the ph interaction
while calculating diagram (b) seen in Fig.~\ref{fig:diagrams}. To that
end, we use the Skyrme interaction as it allows for easy computation
of the RPA diagrams. For consistency, the same Skyrme interaction is
also used in the calculation of the effective mass.  We compare
different parameterizations of the Saclay-Lyon family and of the more
recent Brussels-Montreal family of Skyrme interactions. For the 3p1h
and 1p3h vertices, we use the $\vlowk$ interaction at a cutoff
$\Lambda = \min(2.5\, k_F, 2 \fmi)$, which is also used as bare pp
interaction. As noted in~\cite{Ramanan2018}, with the variable cutoff
one can correctly account for the screening at low densities (GMB
limit).

With the BSk and SLy families of interaction, there is model
dependence already at the level of the bare $T_c$ due to differences
in the effective mass. Surprisingly, the model dependence is reduced
for the screened $T_c$, computed with the full RPA. Comparing the
full RPA results with those of the Landau approximation, one finds
that the Landau approximation is only valid at very low densities
($k_F\lesssim 0.4\fmi$, corresponding to $\rho\lesssim
0.002\fm^{-3}$). At higher densities, screening is stronger i.e.,
$T_c$ is lower, with the full RPA. In particular, the anti-screening
observed with the Landau approximation in~\cite{Ramanan2018} is
completely lost, except with SLy4 in a small range of densities. In
addition, one observes a correlation between the repulsion in the
Landau parameter $G_0$ and the extent of screening, i.e., the more
repulsive $G_0$ gets, the less screened is the dressed
interaction. Qualitatively, this is easily understood in the Landau
approximation, but it is interesting that this correlation is still
present with the full RPA.

In this paper, we also include the correlations due to the preformed
pairs via the NSR approach. While the effect from the preformed pairs
is less drastic compared to the inclusion of medium polarization, we
note that the correlations within NSR show up at low densities, where
screening is strong even in the Landau approximation.  As a result,
the NSR effect of correlations on the density is weaker once screening
is included.

While the transition temperatures at low density (say, up to
$k_F\simeq 0.9\fmi$, corresponding to $\sim 1/7$ of saturation
density) are relatively robust, the high-density region is very
sensitive to the approximation used. As noted here, momentum
dependence of the ph interaction (and probably also of the effective
mass, which is constant in the case of Skyrme interactions) seems to
be a crucial ingredient in the extent to which the bare interaction is
dressed. Perhaps techniques that allow for the building of the
correlations from the medium, such as, in-medium similarity
renormalization group might help in understanding the high-density
region. Such a direction is crucial especially for a realistic
description of pairing in the triplet channel.

\begin{acknowledgments}
We would like to thank D. Davesne and A. Pastore for useful
discussions about RPA response functions.
\end{acknowledgments}

\appendix
\section{Particle-hole interaction from Skyrme forces of the BSk family}
\label{app:Skyrme}
The energy functional corresponding to the generalized Skyrme force
given in Eqs.~(1) and (5) of \Ref{Chamel2009} can be found in
Appendix~A of \Ref{Pastore2014}. Let us write it down (correcting a
typo in the $C^{\nabla s}$ term) for the case of pure neutron matter:
\begin{multline}
\mathcal{E}_{\text{Sk}} = C^\rho \rho^2
  + C^{\tau} (\rho\tau-\jv^2)
  + C^{\nabla\rho} (\nablav \rho)^2
  + C^{s} \sv^2\\
  + C^{T} (\sv\cdot\Tv-\JJ^2)
  + C^{\laplace s} \sv\cdot\laplace\sv
  + C^{\nabla s} (\nablav\otimes\sv)^2\\
  + C^{\nabla J} (\rho\nablav\cdot\Jv+\sv\cdot\nablav\times\jv)\,.
  \label{eq:BSkfunctional}
\end{multline}
For the definitions of the quantities $\rho$, $\tau$, $\jv$, $\sv$,
$\Tv$, $\JJ$, and $\Jv$, see \Ref{Bender2003}. Since we are dealing
with pure neutron matter, the coefficients $C^i$ are the sum of
isoscalar and isovector coefficients: $C^i = C^i_0+C^i_1$. While in
standard Skyrme interactions only $C^\rho$ and $C^{s}$ are density
dependent, this is now the case for all the $C^i$ except $C^{\nabla
  J}$. In \Eq{eq:BSkfunctional}, we have combined the
$\rho\laplace\rho$ term of \Ref{Pastore2014} with the
$(\nablav\rho)^2$ term using integration by parts. Hence, in terms of
the coefficients of \Ref{Pastore2014}, our coefficient
$C^{\nabla\rho}$ corresponds to $C^{\nabla\rho}-(\rho
C^{\laplace\rho})'$, where $X' = dX/d\rho$. In contrast, the
$\sv\laplace\sv$ term cannot be completely absorbed in the
$(\nablav\otimes\sv)^2$ term (or vice versa) because of the density
dependence of $C^{\laplace s}$ and $C^{\nabla s}$, and we therefore
keep both terms. Using the abbreviations
\begin{align}
  \ttil_0 &= t_0(1-x_0)\,, &
  \ttil_1 &= t_1(1-x_1)\,, &
  \ttil_2 &= t_2(1+x_2)\,,\nonumber\\
  \ttil_3 &= t_3(1-x_3)\,, &
  \ttil_4 &= t_4(1-x_4)\,, &
  \ttil_5 &= t_5(1+x_5)\,,
\end{align}
the coefficients $C^{i}$ can be written as
\begin{align}
  C^\rho &= \tfrac{1}{4}\ttil_0+\tfrac{1}{24}\ttil_3\rho^\alpha\,,\nonumber\\
  C^\tau &= \tfrac{1}{8}(\ttil_1+3\ttil_2+\ttil_4\rho^\beta
    +3\ttil_5\rho^\gamma)\,,\nonumber\\
  C^{\nabla\rho} &= \tfrac{3}{32}(\ttil_1-\ttil_2
    +(1+\tfrac{2}{3}\beta)\ttil_4\rho^\beta-\ttil_5\rho^\gamma)\,,\nonumber\\
  C^s &= -\tfrac{1}{4}\ttil_0-\tfrac{1}{24}\ttil_3\rho^\alpha\,,\nonumber\\
  C^{T} &= \eta_J\tfrac{1}{8}(-\ttil_1+\ttil_2-\ttil_4\rho^\beta
    +\ttil_5\rho^\gamma)\,,\nonumber\\
  C^{\laplace s} &= \tfrac{1}{32}(3\ttil_1+\ttil_2+2\ttil_4\rho^\beta)\,,\nonumber\\
  C^{\nabla s} &= -\tfrac{1}{32}(\ttil_4\rho^\beta+\ttil_5\rho^\gamma)\,,\nonumber\\
  C^{\nabla J} &= -W_0\,.
\label{eq:Cfromt}
\end{align}
For interactions which were fitted without the $\JJ^2$ terms (e.g.,
SLy4 \cite{Chabanat1998}, BSk19-21 \cite{Goriely2010}) one should use
$\eta_J=0$, otherwise $\eta_J=1$ (e.g., for SLy5
\cite{Chabanat1998}).

Inserting the functional (\ref{eq:BSkfunctional}) into
\Eq{eq:functionalderivative}, one obtains the particle-hole
interaction. In the case of the BSk interactions, the additional
density dependence of $C^\tau$ leads to a slightly more general form
than \Eq{eq:Skyrmeforce}. To be specific, instead of one coefficient
$v^0_{2}$, one needs now two independent coefficients $v^0_{2}$ and
$v^0_{3}$ analogous to \Eq{eq:v1-18}:
\begin{gather}
  v^0_{1}(q) = (\rho^2 C^{\rho})''+(\rho C^{\tau})''\tau
    + \big[2C^{\nabla\rho}-\tfrac{1}{2}(\rho C^{\tau})'\big] q^2\,,\nonumber\\
  v^0_{2} = (\rho C^{\tau})'\,,\quad
  v^0_{3} = -2C^{\tau}\,,\nonumber\\
  v^0_{4}(q) = 2C^{s}-2\big(C^{\laplace s}-C^{\nabla s}+\tfrac{1}{4}C^{sT}\big) q^2\,,
  \nonumber\\
  v^0_{5} = C^{sT}\,,\quad v^0_{6} = -2C^{sT}\,,\quad
  v^0_{8} = v^0_{9} = C^{\nabla J}\,,\label{eq:v0fromC}
\end{gather}
with $\tau = 3\rho k_F^2/5$. All other $v^0_{i}$ are zero. The
expressions for the $v^0_{i}$ in terms of the Skyrme parameters
$\ttil_i$ are readily obtained by inserting \Eqs{eq:Cfromt} into
\Eqs{eq:v0fromC}. They can also be obtained from the $\bar{W}^{(S)}_i$
of \Ref{Pastore2014}: $v^0_{1} = \bar{W}^{(0)}_1/2-\bar{W}^{(0)}_3
q^2/4$, $v^0_{2} = \bar{W}^{(0)}_2/2$, $v^0_{3} =
\bar{W}^{(0)}_3-\bar{W}^{(0)}_2$, $v^0_{4} =
\bar{W}^{(1)}_1/2-\bar{W}^{(1)}_3 q^2/4$, $v^0_{5} = \bar{W}^{(1)}_2/2$,
and $v^0_{6} = \bar{W}^{(1)}_3-\bar{W}^{(1)}_2$.

\section{Fermi liquid parameters}
\label{app:FL}
From the ph interaction, it is straight forward to get the
effective mass $m^*$ and the lowest-order Landau parameters:
\begin{align}
  \frac{1}{m^*} &= \frac{1}{m}+2 \rho C^{\tau}\,.\label{eq:mstar}\\
  f_0 &= v^0_{1}(q=0)+2v^0_{2} k_F^2\nonumber\\
    &= (\rho^2C^{\rho})''
    +\big[\tfrac{3}{5}\rho(\rho C^{\tau})''+2(\rho C^{\tau})'\big]k_F^2\,,
    \label{eq:f0}\\
    g_0 &= v^0_{4}(q=0)+2v^0_{5} k_F^2
    = 2C^{s}+2C^{sT}k_F^2\,.\label{eq:g0}
\end{align}
The dimensionless Landau parameters $F_0$ and $G_0$ shown in
\Fig{fig:FL} are defined as $F_0 = N_0f_0$ and $G_0 = N_0g_0$. Our
expressions (\ref{eq:mstar}), (\ref{eq:f0}), and (\ref{eq:g0}) for the
Fermi-liquid parameters coincide with those given in Eqs.~(B3), (B1a),
and (B1c) of \Ref{Goriely2010}.

\section{Generalized Lindhard functions}
\label{app:Lindhard}
The generalized Lindhard functions $\Pi_n$ can be computed
analytically. It is convenient to write them as $\Pi_n =
N_0k_F^n\Pitil_n$, where $\Pitil_n$ are dimensionless functions of
$\qtil = q/k_F$:
\begin{align}
  \Pitil_0 &= -\frac{1}{2}-\frac{4-\qtil^2}{8\qtil}
    \ln\Big|\frac{2+\qtil}{2-\qtil}\Big|\,,
    \label{eq:Pi0}\\
  \Pitil_2 &= -\frac{12-\qtil^2}{16}-\frac{(4-\qtil^2)^2}{64\qtil}
    \ln\Big|\frac{2+\qtil}{2-\qtil}\Big|\,,\label{eq:Pi2}\\
  \Pitil_{2L} &= -\frac{1}{3}\,,\label{eq:Pi2L}\\
  \Pitil_4 &= -\frac{240-8\qtil^2+3\qtil^4}{288}
  -\frac{(4-\qtil^2)^3}{384\qtil}
    \ln\Big|\frac{2+\qtil}{2-\qtil}\Big|\,.\label{eq:Pi4}
\end{align}
The function $\Pi_{2T}$ can be obtained from $\Pi_2$ and $\Pi_{2L}$
according to \Eq{eq:Pi2T}.

Notice that our functions $\Pi_n$ are defined differently from
those in \Refs{Garcia1992,Pastore2015}.

\section{Solution of the RPA equation (\ref{eq:RPA})}
\label{app:RPA}
The Skyrme interaction $\calV^0$ can be written in a form analogous to
\Eq{eq:v1-18} (similar to \Eq{eq:Skyrmeforce} but generalized to
$v^0_{3}\neq -2v^0_{2}$ in the case of BSk interactions, see
Appendix~\ref{app:Skyrme}), with the non-vanishing coefficients
$v^0_{i}$ given in \Eqs{eq:v0fromC} and all other $v^0_{i}=0$. Inserting
this and \Eq{eq:v1-18} into \Eq{eq:RPA}, one gets
\begin{widetext}
{\allowdisplaybreaks[1]
 \begin{align}
  \calV_{21} =& \,\calV^0_{21}\nonumber\\
  &+ \{v^0_{1}\, \Pi_0 + v^0_{2}\,[p_2^2\Pi_0+\Pi_2]
    - v^0_{8}\, i\qv\dotp\pv_2\crossp\sigv_2\,\Pi_0\} v_{1}\nonumber\\
  &+ \{v^0_{1}\, [p_1^2\Pi_0+\Pi_2]
    + v^0_{2}\, [p_1^2p_2^2\Pi_0+(p_1^2+p_2^2)\Pi_2+\Pi_4]
    - v^0_{8}\, i\qv\dotp\pv_2\crossp\sigv_2\,[p_1^2\Pi_0+\Pi_2]\} v_{2}
    \nonumber\\
  &+ \{v^0_{3}\, [\pv_1\dotp\pv_2\,\Pi_{2T}
    +\pv_1\dotp\qv\,\pv_2\dotp\qv\,(\Pi_{2L}-\Pi_{2T})/q^2]
    + v^0_{9}\, i\qv\dotp\pv_1\crossp\sigv_2\Pi_{2T}\} v_{3}\nonumber\\
  &+ \{v^0_{4}\, \sigv_1\dotp\sigv_2\,\Pi_0
    + v^0_{5}\, \sigv_1\dotp\sigv_2\,[p_2^2\Pi_0+\Pi_2]
    - v^0_{9}\, i\qv\dotp\pv_2\crossp\sigv_1\,\Pi_0\} v_{4}\nonumber\\
  &+ \{v^0_{4}\, \sigv_1\dotp\sigv_2\,[p_1^2\Pi_0+\Pi_2]
    + v^0_{5}\, \sigv_1\dotp\sigv_2\,[p_1^2p_2^2\Pi_0+(p_1^2+p_2^2)\Pi_2+\Pi_4]
    - v^0_{9}\, i\qv\dotp\pv_2\crossp\sigv_1\,[p_1^2\Pi_0+\Pi_2]\} v_{5}
    \nonumber\\
  &+ \{v^0_{6}\, \sigv_1\dotp\sigv_2\,[\pv_1\dotp\pv_2\,\Pi_{2T}
       + \pv_1\dotp\qv\,\pv_2\dotp\qv\,(\Pi_{2L}-\Pi_{2T})/q^2]
    + v^0_{8}\, i\qv\dotp\pv_1\crossp\sigv_1\,\Pi_{2T}\} v_{6}\nonumber\\
  &+ \{v^0_{4}\, \sigv_1\dotp\qv\,\sigv_2\dotp\qv \Pi_0
    + v^0_{5}\, \sigv_1\dotp\qv\,\sigv_2\dotp\qv\,[p_2^2\Pi_0+\Pi_2]\} v_{7}
    \nonumber\\
  &+ \{v^0_{1}\, i\qv\dotp\pv_1\crossp\sigv_1\,\Pi_0
    + v^0_{2}\, i\qv\dotp\pv_1\crossp\sigv_1\,[p_2^2\Pi_0+\Pi_2]
    - v^0_{6}\, i\qv\dotp\pv_2\crossp\sigv_2\Pi_{2T}
    + v^0_{8} \,[\qv\dotp\pv_1\crossp\sigv_1\,\qv\dotp\pv_2\crossp\sigv_2\,\Pi_0
      +2q^2\Pi_{2T}]\} v_{8}\nonumber\\
  &+ \{-v^0_{3}\, i\qv\dotp\pv_2\crossp\sigv_1\,\Pi_{2T}
    + v^0_{4}\, i\qv\dotp\pv_1\crossp\sigv_2\,\Pi_0
    + v^0_{5}\, i\qv\dotp\pv_1\crossp\sigv_2\,[p_2^2\Pi_0+\Pi_2]\nonumber\\*
  &\phantom{+\{}
    + v^0_{9}\, [(\pv_1\dotp\pv_2\,q^2-\pv_1\dotp\qv\,\pv_2\dotp\qv)\Pi_0
      +(\sigv_1\dotp\sigv_2\,q^2
        -\sigv_1\dotp\qv\,\sigv_2\dotp\qv)\Pi_{2T}]\} v_{9}\nonumber\\
  &+ \{v^0_{1}\, p_1^2\Pi_2
    + v^0_{2}\, p_1^2\,[p_2^2\Pi_2+\Pi_4]
    - v^0_{8}\, i\qv\dotp\pv_2\crossp\sigv_2\, p_1^2 \Pi_2\} v_{10}\nonumber\\
  &+ v^0_{3}\, \pv_1\dotp\qv\,\pv_2\dotp\qv\,\Pi_{2L} v_{11}\nonumber\\
  &+ \{v^0_{4}\, \sigv_1\dotp\sigv_2\,p_1^2\Pi_2
    + v^0_{5}\, \sigv_1\dotp\sigv_2\,p_1^2\,[p_2^2\Pi_2+\Pi_4]
    - v^0_{9}\, i\qv\dotp\pv_2\crossp\sigv_1\,p_1^2\Pi_2\} v_{12}\nonumber\\
  &+ v^0_{6}\, \sigv_1\dotp\sigv_2\,\pv_1\dotp\qv\,\pv_2\dotp\qv\,\Pi_{2L} v_{13}
    \nonumber\\
  &+ \{v^0_{4}\, \sigv_1\dotp\qv\,\sigv_2\dotp\qv\,[p_1^2\Pi_0+\Pi_2]
    + v^0_{5}\, \sigv_1\dotp\qv\,\sigv_2\dotp\qv\,
      [p_1^2p_2^2\Pi_0+(p_1^2+p_2^2)\Pi_2+\Pi_4]\} v_{14}\nonumber\\
  &+ \{v^0_{4}\, \sigv_1\dotp\qv\,\sigv_2\dotp\qv\,p_1^2\Pi_2
    + v^0_{5}\, \sigv_1\dotp\qv\,\sigv_2\dotp\qv\,p_1^2\,
      [p_2^2\Pi_2+\Pi_4]\} v_{15}\nonumber\\
  &+ \{v^0_{1}\, i\qv\dotp\pv_1\crossp\sigv_1\,\Pi_2
    + v^0_{2}\, i\qv\dotp\pv_1\crossp\sigv_1\,[p_2^2 \Pi_2+\Pi_4]
    - v^0_{6}\, i\qv\dotp\pv_2\crossp\sigv_2\,p_1^2\Pi_{2T}\nonumber\\*
  &\phantom{+\{}
    + v^0_{8}\, [\qv\dotp\pv_1\crossp\sigv_1\,\qv\dotp\pv_2\crossp\sigv_2\,\Pi_2
      +2 q^2 p_1^2\Pi_{2T}]\} v_{16}\nonumber\\
  &+ \{-v^0_{3}\, i\qv\dotp\pv_2\crossp\sigv_1\, p_1^2 \Pi_{2T}
    + v^0_{4}\, i\qv\dotp\pv_1\crossp\sigv_2\, \Pi_2
    + v^0_{5}\, i\qv\dotp\pv_1\crossp\sigv_2\, [p_2^2\Pi_2+\Pi_4]\nonumber\\*
  &\phantom{+\{}
    + v^0_{9}\, [(\pv_1\dotp\pv_2\, q^2-\pv_1\dotp\qv\,\pv_2\dotp\qv)\Pi_2
      +(\sigv_1\dotp\sigv_2\, q^2-\sigv_1\dotp\qv\,\sigv_2\dotp\qv)
        p_1^2\Pi_{2T}]\} v_{17}\nonumber\\
  &+ \{v^0_{6}\, \qv\dotp\pv_1\crossp\sigv_1\,\qv\dotp\pv_2\crossp\sigv_2 \Pi_{2T}
    + v^0_{8}\, i\qv\dotp\pv_1\crossp\sigv_1\, 2q^2\Pi_{2T}\} v_{18}\,.
 \end{align}
}
\end{widetext}

By collecting the coefficients of the different operators, we obtain a
system of linear equations for the unknown $v_{i}$, of the form
(\ref{eq:RPAmatrixform}).

Notice that for some of the $v_{i}$, the equations are not unique. For
instance, the equation for $v_{2}$ can be obtained from the
coefficients of $p_1^2$ or $p_2^2$. However, the final result is
independent of this choice because the equality of the coefficients
follows from the hermiticity of $V_{12} = V^*_{21}$. Here, we choose
the equations for $v_{2}$, $v_{5}$, $v_{8}$, $v_{9}$, $v_{14}$,
$v_{16}$, and $v_{17}$ that are obtained from the coefficients of
$p_2^2$, $\sigv_1\dotp\sigv_2\, p_2^2$,
$-i\qv\dotp\pv_2\crossp\sigv_2$, $-i\qv\dotp\pv_2\crossp\sigv_1$,
$\sigv_1\dotp\qv\,\sigv_2\dotp\qv\,p_2^2$,
$i\qv\dotp\pv_1\crossp\sigv_1\,p_2^2$, and
$i\qv\dotp\pv_1\crossp\sigv_2\,p_2^2$, respectively.

With this choice, the non-vanishing matrix elements in
\Eq{eq:RPAmatrixform} are:
{\allowdisplaybreaks[1]
\begin{gather}
  A_{1,1} = v^0_{1} \Pi_0 + v^0_{2} \Pi_2\,,\quad
  A_{1,2} = v^0_{1} \Pi_2 + v^0_{2} \Pi_4\,,\nonumber\\
  A_{1,8} = 2 v^0_{8} q^2 \Pi_{2T}\,,\quad
  A_{2,1} = v^0_{2} \Pi_0\,,\quad
  A_{2,2} = v^0_{2} \Pi_2\,,\nonumber\\
  A_{3,3} = v^0_{3} \Pi_{2T}\,,\quad
  A_{3,9} = v^0_{9} q^2 \Pi_0\,,\quad
  A_{3,17} = v^0_{9} q^2 \Pi_2\,,\nonumber\\
  A_{4,4} = v^0_{4} \Pi_0 + v^0_{5} \Pi_2\,,\quad
  A_{4,5} = v^0_{4} \Pi_2 + v^0_{5} \Pi_4\,,\nonumber\\
  A_{4,9} = v^0_{9} q^2\Pi_{2T}\,,\quad
  A_{5,4} = v^0_{5} \Pi_0\,,\quad
  A_{5,5} = v^0_{5} \Pi_2\,,\nonumber\\
  A_{6,6} = v^{0}_{6} \Pi_{2T}\,,\quad
  A_{7,7} = v^0_{4} \Pi_0 + v^0_{5} \Pi_2\,,\nonumber\\
  A_{7,9} = -v^0_{9} \Pi_{2T}\,,\quad
  A_{7,14} = v^0_{4} \Pi_2 + v^0_{5} \Pi_4\,,\nonumber\\
  A_{8,1} = v^0_{8} \Pi_0\,,\quad
  A_{8,2} = v^0_{8} \Pi_2\,,\quad
  A_{8,8} = v^0_{6} \Pi_{2T}\,,\nonumber\\
  A_{9,4} = v^0_{9} \Pi_0\,,\quad
  A_{9,5} = v^0_{9} \Pi_2\,,\quad
  A_{9,9} = v^0_{3} \Pi_{2T}\,,\nonumber\\
  A_{10,2} = v^0_{2} \Pi_0\,,\quad
  A_{10,10} = v^0_{2} \Pi_2\,,\quad\nonumber\\
  A_{11,3} = v^0_{3} (\Pi_{2L}-\Pi_{2T})/q^2\,,\quad
  A_{11,9} = -v^0_{9} \Pi_0\,,\nonumber\\
  A_{11,11} = v^0_{3} \Pi_{2L}\,,\quad
  A_{11,17} = -v^0_{9} \Pi_2\,,\quad
  A_{12,5} = v^0_{5} \Pi_0\,,\nonumber\\
  A_{12,12} = v^0_{5} \Pi_2\,,\quad
  A_{13,6} = v^0_{6} (\Pi_{2L}-\Pi_{2T})/q^2\,,\nonumber\\
  A_{13,13} = v^0_{6} \Pi_{2L}\,,\quad
  A_{14,7} = v^0_{5} \Pi_0\,,\quad
  A_{14,14} = v^0_{5} \Pi_2\,,\nonumber\\
  A_{15,14} = v^0_{5} \Pi_0\,,\quad
  A_{15,15} = v^0_{5} \Pi_2\,,\quad
  A_{16,8} = v^0_{2} \Pi_0\,,\nonumber\\
  A_{16,16} = v^0_{2} \Pi_2\,,\quad
  A_{17,9} = v^0_{5} \Pi_0\,,\quad
  A_{17,17} = v^0_{5} \Pi_2\,,\nonumber\\
  A_{18,8} = v^0_{8} \Pi_0\,,\quad
  A_{18,16} = v^0_{8} \Pi_2\,,\quad
  A_{18,18} = v^0_{6} \Pi_{2T}\,.
\end{gather}
This system of equations has actually two decoupled blocks
corresponding to the indices 1, 2, 6, 8, 10, 13, 16, 18, which have
products of two time-even operators, and to the indices 3, 4, 5, 7, 9,
11, 12, 14, 15, 17, which have products of two time-odd operators.

The solution for the time-even operators reads:
\begin{gather}
  v_{1} = \frac{\phi_a-1}{\Pi_0}\,,\quad
  v_{2} = v^0_{2} \phi_2 \phi_a\,,\quad
  v_{6} = v^0_{6} \phi_{6T}\,,\nonumber\\
  v_{8} = v^0_{8} \phi_2 \phi_{6T} \phi_a\,,\quad
  v_{10} = (v^0_{2})^2 \Pi_0 \phi_2^2 \phi_a\,,\nonumber\\
  v_{13} = \frac{(v^0_{6})^2 (\Pi_{2L}-\Pi_{2T}) \phi_{6L} \phi_{6T}}{q^2}\nonumber\\
  v_{16} = v^0_{2} v^0_{8} \Pi_0 \phi_2^2 \phi_{6T} \phi_a\,,\quad
  v_{18} = (v^0_{8})^2 \Pi_0 \phi_2^2 \phi_{6T}^2 \phi_a\,,
\end{gather}
and the solution for the time-odd operators:
\begin{gather}
  v_{3} = \frac{\phi_{3T}\phi_c-1}{\Pi_{2T}}\,,\quad
  v_{4} = \frac{\phi_b\phi_c-\phi_5^2}{\Pi_0\phi_5^2}\,,\quad
  v_{5} = \frac{v^0_{5}\phi_b\phi_c}{\phi_5}\,,\nonumber \\
  v_{7} = -\frac{(v^0_{9})^2\Pi_{2T}\phi_{3T}\phi_b^2\phi_c}{\phi_5^2}\,,\quad
  v_{9} = \frac{v^0_{9}\phi_{3T}\phi_b\phi_c}{\phi_5}\,,\nonumber \\
  v_{11} = \frac{\phi_{3L}[1-\phi_c-(1-\phi_{3T}\phi_c)v^0_{3}(\Pi_{2L}-\Pi_{2T})]}
    {q^2\Pi_{2T}}\,,\nonumber \\
  v_{12} = (v^0_{5})^2\Pi_0\phi_b\phi_c \,,\quad
  v_{14} = -\frac{v^0_{5}(v^0_{9})^2\Pi_0\Pi_{2T}\phi_{3T}\phi_b^2\phi_c}{\phi_5}\,,
  \nonumber \\
  v_{15} = -(v^0_{5} v^0_{9})^2\Pi_0^2\Pi_{2T}\phi_{3T}\phi_b^2\phi_c\,,\nonumber \\
  v_{17} = v^0_{5}v^0_{9}\Pi_0\phi_{3T}\phi_b\phi_c\,,
\end{gather}
where the following abbreviations have been used:
\begin{gather}
  \phi_2 = \frac{1}{1 - v^0_{2} \Pi_2}\,,\quad
  \phi_{3L,T} = \frac{1}{1 - v^0_{3} \Pi_{2L,T}}\,,\nonumber\\
  \phi_5 = \frac{1}{1 - v^0_{5} \Pi_2}\,,\quad
  \phi_{6L,T} = \frac{1}{1 - v^0_{6} \Pi_{2L,T}}\,,\nonumber\\
  \phi_a = \frac{1}{1 - \Pi_0 [v^0_{1} + (v^0_{2})^2 \Pi_4 + 
      2 q^2 (v^0_{8})^2 \Pi_{2T} \phi_{6T}]\phi_2^2}\nonumber\\
  \phi_b = \frac{1}{1 - v^0_{4} \Pi_0 - 2 v^0_{5} \Pi_2 + 
    (v^0_{5})^2 (\Pi_2^2 - \Pi_0 \Pi_4)}\,,\nonumber\\
  \phi_c = \frac{1}{1 - q^2 (v^0_{9})^2 \Pi_0 \Pi_{2T} \phi_{3T} \phi_b}\,.
\end{gather}}
\section{Correlated density (NSR correction)}
\label{app:NSR}
In this appendix we summarize the formulas needed for the calculation
of the NSR correction. For their derivation and more details, see
\Refs{Ramanan2013,Ramanan2018}. For a given chemical potential $\mu$,
the neutron density $\rho$ is written as a sum of three terms,
\begin{equation}
  \rho = \rho_0+\rho_{\corr,1}+\rho_{\corr,2}.
\end{equation}
The first term, $\rho_0$, is the uncorrelated density,
\begin{equation}
  \rho_0 = 2 \sum_{\vek{p}} f(\xi_p)\,,
\end{equation}	
where $f(\xi_p) = 1/(e^{\xi_p/T}+1)$ is the Fermi-Dirac distribution
function, $\xi_p = \epsilon_p-\mu$, and the factor of
$2$ arises due to the spin degeneracy. The other terms are the
correlated density calculated to first order in the single-particle
self-energy $\Sigma$. The term $\rho_{\corr,1}$ corresponds to the
original NSR correction \cite{Nozieres1985} and is given by
\begin{equation}
  \rho_{\corr,1} = 2 T \sum_{\pv,\omega_n} [\Gtemp_0(\pv, \omega_n)]^2
  \Sigma(\pv, i\omega_n)\,,
 \label{eq:rhocorr1}
\end{equation}
where $\omega_n$ are the fermionic Matsubara frequencies and $\Gtemp_0
= 1/(i\omega_n-\xi_{{p}})$ is the uncorrelated single-particle Green's
function in the imaginary-time formalism~\cite{FetterWalecka}.
Calculating $\Sigma({\vek{p}}, i\omega_n)$ within the ladder
approximation and performing a couple of steps detailed in
\cite{Ramanan2013}, one finally obtains the expression
\begin{equation}
  \rho_{\corr,1} = -\frac{\partial}{\partial \mu}
  \sum_{\Kv,\nu}\int\! \frac{d\omega}{\pi}\, g(\omega)
  \im \log[1 - \eta_\nu(K, \omega)]\,.
\end{equation}
Here, $g(\omega) = 1/(e^{\beta \omega} - 1)$ is the Bose function and
$\eta_\nu$ are the complex eigenvalues of
\begin{equation}
  V\overline{G}^{(2)} = V(k,k')\frac{\overline{Q}(K,k')}
  {\omega-\frac{K^2}{4m^*}-\frac{k^{\prime\,2}}{m^*}+2\mu}\,,
\end{equation}
where $\overline{Q}$ denotes the angle average (since we consider only
the $s$ wave) of the Pauli blocking factor $Q(\Kv,\kv) =
1-f(\xi_{\frac{\Kv}{2}-{\kv}})-f(\xi_{\frac{\Kv}{2}+{\kv}})$
where $\Kv$ is the total momentum of the pair. In principle, the
  screening correction to $V(k,k')$ should also depend on
  $K$, but as shown in \cite{Ramanan2018} this dependence is weak and
  we neglect it.

The last term, $\rho_{\corr,2}$ is absent in the original NSR
approach. It takes into account that $\Gtemp_0$ includes already the
in-medium quasiparticle energy $\xi_p$ which therefore must not be
shifted by the self-energy \cite{Zimmermann1985,Jin2010}. Accordingly,
one has to subtract from the self-energy in \Eq{eq:rhocorr1} its
on-shell value $\re \Sigma(\vek{p},\xi_p)$, which gives
\begin{equation}
  \rho_{\corr,2} = -2 T \sum_{\vek{p},\omega_n} [\Gtemp_0(\vek{p},
    \omega_n)]^2 \re \Sigma(\vek{p},\xi_p)\,.
 \label{eq:rhocorr2}
\end{equation}
As in \cite{Ramanan2013,Ramanan2018}, we approximate
$\Sigma(\vek{p},\xi_p)$ by the first-order (Hartree-Fock) self-energy
and finally arrive at the following correction:
\begin{equation}
  \rho_{\corr,2} = 4\pi\frac{\partial}{\partial \mu}
  \sum_{\Kv,\kv} g\big(\tfrac{K^2}{4m^*} + \tfrac{k^2}{m^*} - 2\mu\big)
    V(k,k)\overline{Q}(K,k)\,.
\end{equation} 


\end{document}